# Turning bad into good: a water-splitting-active hole transporting material to preserve the performance of perovskite solar cells in humid environments †

Min Kim,[1,#] Antonio Alfano,[1,2,#] Giovanni Perotto,[3] Michele Serri,[4] Nicola Dengo,[5] Alessandro Mezzetti,[1] Silvia Gross,[5,6] Mirko Prato,[7] Marco Salerno,[7] Roberto Sorrentino,[1] Gaudenzio Meneghesso,[6,8] Fabio Di Fonzo,[1] Annamaria Petrozza,[1] Teresa Gatti[9]* Francesco Lamberti[5,6]*

[1] Center for Nano Science and Technology, Istituto Italiano di Tecnologia, Via Pascoli 70/3, 20133 Milano, Italy

[2] Department of Physics, Politecnico di Milano, P.zza L. da Vinci 32, 20133 Milano, Italy

[3] Nanophysics Department, Smart Materials, Istituto Italiano di Tecnologia, Via Morego 30, 16163 Genova, Italy

[4] Istituto Italiano di Tecnologia – Graphene Labs, via Morego 30, 16163 Genova, Italy

[5] Department of Chemical Sciences, University of Padova and INSTM UdR Padova, Via Marzolo 1, 35131 Padova, Italy

[6] Center "Giorgio Levi Cases" for Energy Economics and Technology, Via Marzolo 9, 35131 Padova, Italy

[7] Materials Characterization Facility, Istituto Italiano di Tecnologia, Via Morego 30, 16163 Genova, Italy

[8] Department of Information Engineering, University of Padova, Via Gradenigo 6/B, 35131 Padova, Italy

[9] Center for Materials Research, Justus Liebig University Giessen, Heinrich Buff Ring 17, 35392 Giessen, Germany

[#] These authors contributed equally

* Correspondence: <u>francesco.lamberti@unipd.it</u>, <u>teresa.gatti@phys.chemie.uni-giessen.de</u>

† Supporting Information (SI) available: further optical, spectroscopical and morphological characterizations are provided.

**Abstract**

Lead halide perovskite-based photoactive layers are nowadays employed for a large number of optoelectronic applications, from solar cells to photodetectors and light-emitting diodes, because of their excellent absorption, emission and charge-transport properties. Unfortunately, their commercialization is still hindered by an intrinsic instability towards classical environmental conditions. Water in particular promotes fast decomposition, leading to a drastic decrease in device performance. An innovative functional approach to overcome this major issue could derive from integrating water-splitting active species within charge extracting layers adjacent to the perovskite photoactive layer, converting incoming water molecules into molecular oxygen and hydrogen before they reach this last one, thus preserving device performance in time. In this work we report for the first time on a perovskite-ancillary layer based on CuSCN nanoplateletes dispersed in a p-type semiconducting polymeric matrix, combining hole extraction/transport properties with good water-oxidation activity, that transforms incoming water molecules and further triggers the *in situ* p-doping of the conjugated polymer by means of the produced dioxygen, further improving transport of photogenerated charges. This composite layer enables the long-term stabilization of a mixed cation lead halide perovskite within a direct solar cell architecture, maintaining a stable performance for 28 days in high-moisture simulated conditions. Our findings demonstrate that the engineering of a hole extraction layer with water-splitting active additives represent a valuable strategy to mitigate the degradation of perovskite solar cells exposed to atmospheric humidity. A similar approach could be employed in the future to improve stabilities of other optoelectronic devices based on water-sensitive species.

**Introduction**

Stability to environmental humidity is the main bottleneck delaying perovskite solar cell (PSC) commercialization.[1,2] It is widely accepted that most of the degradation pathways in photoactive perovskite layers come from water permeated within photovoltaic devices, that irreversibly depresses their figures of merit. Water quickly promotes the transformation of the perovskite salts into their precursors[3] and the process is even faster under light illumination.[4] Silicon solar cells do not suffer from this problem and thus they are steps ahead in the photovoltaic market.

In order to avoid moisture permeation, PSC can be robustly engineered with the application of hydrophobic ancillary layers or with complex device encapsulation.[5] In this way, examples of PSC working for several thousands hours have been reported in the literature.[6,7,8] Nevertheless, despite all the big efforts coming from the scientific community to avoid penetration of water up to the photoactive halide perovskite layers, this major issue still remains extremely relevant for determining device lifetimes and novel disrupting approaches are required to proceed further towards PSC commercialization.

A possible solution could derive from the incorporation of materials having water-splitting (WS) activity within one of the layers adjacent to the perovskite film in a solar cell multilayer architecture. For example, in a direct architecture PSC, the hole transporting layer could be engineered to act as charge extractor/transporter layer and as a functional scavenger for moisture at the same time, hindering water penetration not only through instrinsic hydrophobicity (a very popular strategy to stabilize PSC nowadays) but by transforming the water molecules into other harmless or even beneficial species.

Photocatalytic WS is the process at the base of artificial photosynthesis,[9,10] allowing the chemical dissociation of water into molecular hydrogen and oxygen. It requires energy (light or bias), a catalyst and water. The mechanism is straightforward:[11] a theoretical energy difference of at least 1.23 eV is required in order to promote water oxidation. Different materials, mainly inorganic semiconductors,[12] are known having WS properties due to their favorable energy alignement with respect to the oxygen/hydrogen evolution potential.

Within these scenario, the essential requirement for realizing a WS-active hole-transporting material (HTM) would be to identify species that combine hole transport with water oxidation activity (2 $H_2O$ → $O_2$ + 4$H^+$ + 4$e^-$). A major candidate for this role can be found in copper thiocyanate (CuSCN), which has been shown to act as an excellent HTM,[13] with high hole-mobility, good energy level matching with perovskites, thermal stability, hydrophobic properties and low production costs. The use of CuSCN as HTM in conjunction with a reduced graphene oxide interfacial thin layer with the top metal electrode, provided a remarkable stabilized power-conversion-efficiency (PCE).[14] At the same time, this material has been used to efficiently assist oxygen evolution in CuSCN/$BiVO_4$ photoelectrodes.[15] On the other hand, CuSCN is not used alone for photocatalytic WS, given the scarce absorption properties in the visible range. Within a PSC anyway, the perovskite layer itself would be responsible for light harvesting and hydrogen evolution, due to the favourable energetics of its conduction band towards the reduction of water.

CuSCN large-scale adoption is still limited by the sub-optimal processing employed for the deposition on top of the perovskite layer, which consists in an un-controlled drop-casting from diethyl sulfide (DES) concentrated solutions. Such a procedure is highly user-dependent and therefore suffers from a potentially scarce reproducibility. In addition, it does not ensure a perfect solvent orthogonality with the underlying perovskite, which is exposed to partial dissolution by DES and consequent damage, ultimately affecting the device performance.[14,16] In this regard, the production of CuSCN in the form of solution-processed nanostructures has not yet been reported to the best of our knowledge, but it might allow the use of more suitable solvents and reproducible deposition techniques.

Here, we propose the use of ligand-free CuSCN nanoplatelets (CuSCN-NP) produced through continuous flow hydrothermal synthesis (CFHS). This water-based synthetic route follows the principles of green chemistry and allows to obtain a wide variety of nanomaterials with applications ranging from biomedicine to printed electronics and photocatalysis.[17] The platelet-like geometry of these inorganic nanomaterials is ideal for use in multilayers, thin films-based devices.

On the other hand, this choice also forces to identify a dispersion phase in which incorporate CuSCN-NP for solution processing, that can work at the same time as an hole transporter, in order to facilitate the overall charge extraction process from the perovskite layer. A polymeric hole-transporter such as poly(3-hexylthiophene) (P3HT), a well-established, inexpensive, organic *p*-type polymer characterized by an outstanding film forming ability and a good matching with the copper(I)-based inorganic semiconductor valence band,[18] appears to be the most suitable choice, although undoped P3HT is not generally applied as HTM when high PCEs are targeted.[19–21] By dispersing CuSCN-NP in P3HT, a novel nanocomposite HTM is produced, that can be defined as CuSCN@P3HT.

In this work, we demonstrate that a perovskite/CuSCN@P3HT interface can be used as a WS-active combination for preventing the moisture-mediated degradation of a PSC and for the concomitant *in-situ p*-doping of the P3HT HTM phase, that further affects favorably the figures of merit of the direct architecture PSC in which it is integrated. This virtuous physico-chemical process ensures the maintenance of a good PCE of more than 9% in PSC stored for one month in a water-saturated atmosphere, making these insights a very promising platform for stimulating future investigations on smart approaches to preserve operation in optoelectronic devices[22] suffering from scarce stability to environmental moisture.

**Results and Discussion**

**Materials Synthesis and Characterization**

The synthesis of surfactant-free CuSCN-NP was performed by CFHS, as reported in detail in the Experimental Section and schematized in Fig. S1 in the Supplementary Information (SI). The CuSCN-NP can be visualized in Fig. 1 and Fig. S2 in the SI. From TEM picture (Fig. 1a) and SEM pictures (Fig. S2a), they appear to have a platelet-like shape. Analysis through dynamic light scattering (DLS) (Fig. 1b) show that particles have a narrow size distribution with an average value of 600 ± 160 nm. TEM shows also the presence of a small fraction of particles that are significantly thinner, as shown in the inset of Fig. 1b and in Fig. S2b,c,d,e. These lamellae have a lateral size in the 50–100 nm range and are significantly more electron-transparent than most of other particles, hinting of a thickness of only few nanometeres.

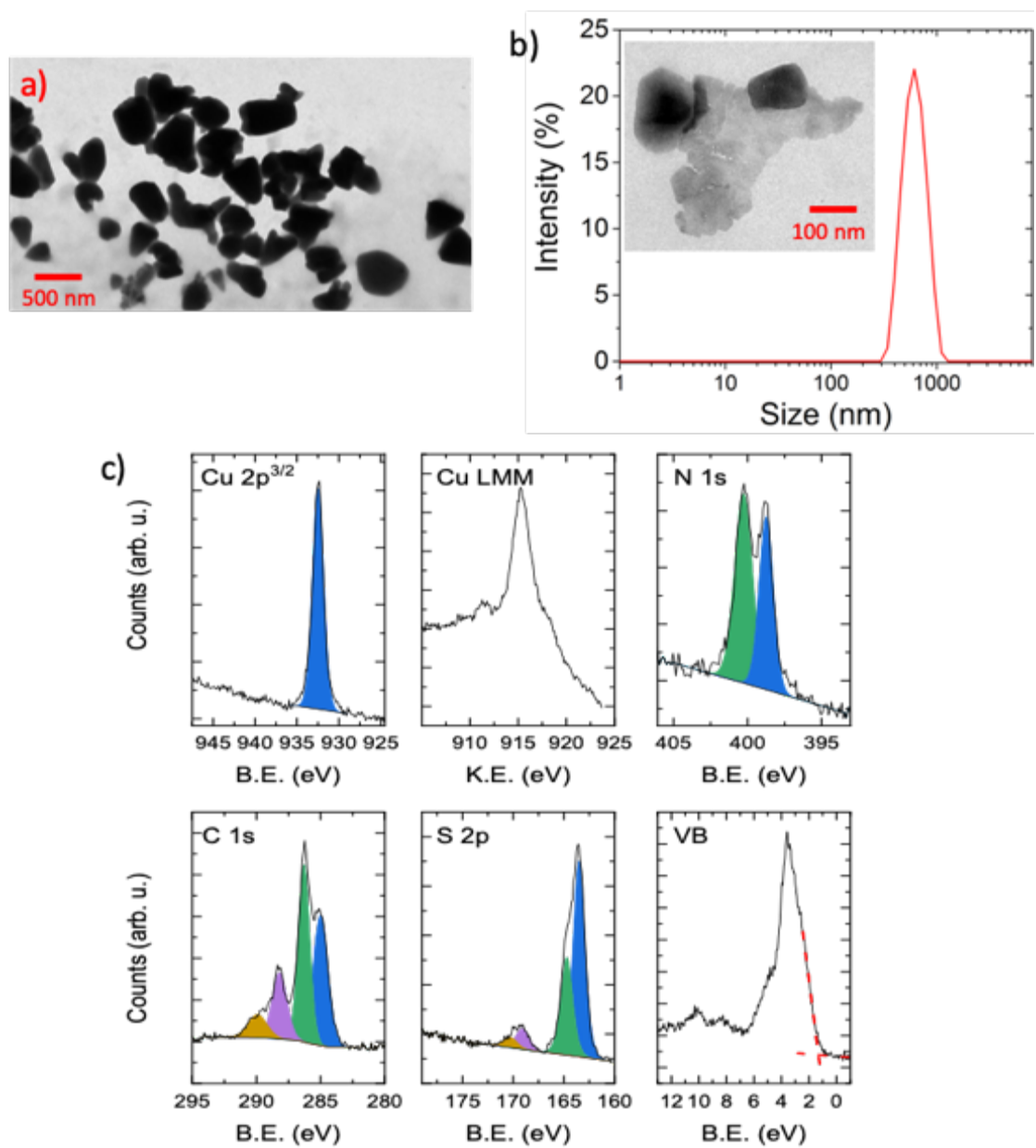

**Figure 1**. Physico-chemical characterization of CuSCN-NP. a) Representative TEM image of CuSCN-NP from water dispersion employing keratin as surfactant and b) size distribution measured by DLS, with TEM image at high magnification in the inset showing the presence of significantly thin platelets within the sample. c) XPS characterization of CuSCN-NP: top left, Cu $2p^{3/2}$ XPS peak with fitted Cu(I)SCN component (blue); top center, Cu LMM Auger spectrum; top right, N 1s XPS spectrum fitted with SCN (blue) and thiourea (green) component; bottom left, C 1s XPS region with aliphatic (blue), SCN (green), thiourea (violet), carboxylate (ochre) components; bottom center, S 2p region with the main doublet assigned to SCN and thiourea ($2p^{3/2}$ blue, $2p^{1/2}$ green), and the satellite doublet corresponding to oxidized sulfur (violet and ochre); bottom right, XPS valence band spectrum with red dashed lines indicating the valence band on-set.

The optical band gap of the CuSCN-NP is slightly higher than the reported one for bulk CuSCN (3.6 eV),[16] as evidenced by the corresponding Tauc Plot (Fig. S3a in the SI), from which an optical direct band gap of 4.1 eV is estimated. The higher value for the band gap can be attributed to the particular aspect ratio of these nanomaterials, with an electronic behavior resembling that of quantum wells.[23] The platelet structure of these nanomaterials is further corroborated by examining the line shape of the Tauc Plot at low energy: this analysis is a well-known methodology for the determination of the optical properties of semiconducting nanoparticles,[24] in particular for revealing the presence of defects, i.e. a continuum of tail states within the particles band-gap.[25] These states (called Urbach states) can be reasonably considered as a measure of the disorder in nanocrystals. An exponential decay of the optical absorption is observed beneath the gap in our CuSCN-NP and an Urbach energy ($E_u$) of about 0.19 eV is estimated by using the relationship:[26]

$$\alpha \approx \exp\left(\frac{E}{E_u}\right)$$

where $\alpha$ is the absorption coefficient (the corresponding Urbach fit is reported in Fig. S3a for the sake of clarity).[25] This value is in agreement with previous literature data on nanostructured CuSCN obtained by electrodeposition in the form of nanowires and 2D platelets.[27]

A more detailed elemental and structural characterization of the material, thorugh X-rays diffraction (XRD) analysis and energy-dispersive X-ray spectroscopy (EDX), shown respectively in Fig. S3b and S3c in the SI, clearly demonstrate that the investigated sample is crystalline and mainly composed of the R-3m trigonal copper thiocyanate phase (β-CuSCN). The presence of preferential orientation of the crystallites suggests that particles diplay a tendency to stack along the [00l] direction (as seen from XRD, Fig. S3b in the SI).[14] This is compatible with the formation of platelets undergoing a preferential growth along the [h k 0] directions. However, the data also indicate the presence of other spurious copper-based phases, evident in the XRD pattern and confirmed by the unproper stoichiometry of Cu:S:C:N (Fig. S3c). In order to deeper investigate the source of this contamination, the structural characterization is completed through X-ray photoelectron spectroscopy (XPS) measurements, shown in Fig. 1c. The Cu 2p3/2 region shows a single component at 932.5 eV, in agreement with values reported for CuSCN (there could be a minor component corresponding to Cu oxides)[28]. The absence of satellite peaks in the binding energy (BE) region 935 — 940 eV suggests that Cu (II) species, in particular Cu(II) oxides and hydroxides, are not present in significant amounts. The attribution of the Cu(I) oxidation state of the metal is confirmed by the Cu LMM Auger spectrum, characterized by a peak at kinetic energy 915.3 eV, and the modified Auger parameter (Cu $2p_{3/2}$ BE + Cu LMM KE) of 1847.7 eV.[29] The contamination is ascribed to thiocyanate ligand and residual thiourea from the synthesis that could not be removed even after the thorough washings, as can be identified by N 1s, C 1s and S 2p peaks at characteristic BE values.[27,30] The N 1s region is characterized by the presence of two peaks at BE 398.7 eV and 400.3 eV, which are assigned to thiocyanate and thiourea respectively. The carbon 1s region exhibits multiple peaks that correspond to the thiocyanate ligand (BE 286.3 eV), alifatic carbon contaminants (BE 285 eV), thiourea (BE 288.2 eV) and carboxylate (BE 289.9 eV) functional groups. Sulfur 2p region is dominated by a doublet corresponding to both SCN and thiourea (S $2p_{3/2}$ BE = 163.5 eV), while a small contribution from oxidized sulfur is present with BE ≈ 169.4 eV. The presence of thiourea detected through XPS suggest a second partial interpration of the increased optical band gap of the CuSCN-NP compared to bulk CuSCN as extrapolated from the Tauc plot in Fig. 1d: a previous literature report on a similar case[31] describes how the band gap of a ZnS thin films resulted to be singificantly affected by the presence of residual thiourea precursor, when a chemical bath deposition process was employed to fabricate it.

A CuSCN@P3HT composite is prepared by blending the ligand-free CuSCN-NP in solid powder form in a P3HT solution, as described in the experimental section. The model solar cells for CuSCN@P3HT HTM integration and test of WS-activity is based on a standard architecture PSC with ITO/SnO$_2$/perovskite/HTM/Au configuration, where tin oxide is the electron transporting layer (ETL). However, before solar cell testing, a clear understanding of the WS process is mandatory.

For this reason, the WS activity of CuSCN@P3HT – in comparison with that of bare P3HT – is investigated by conducting linear sweep voltammetry (LSV) measurements in a 2-electrodes configuration, employing ITO/SnO$_2$ and HTM/Au/ITO partial devices as the two electrodes. Trivially, being impossibile to place it directly into the aqueous buffer solution for the WS measurements due to its intrinsic instability, the perovskite photoactive layer is removed (essentially dividing the solar cell into the two devices used as electrodes) and the photovoltage is artificially provided by the potentiostat, as the LSV measurements are performed in dark. Experimental details on the preparation of the system and on the electrochemical measurements are given in the experimental section. To determine the onset potential for the WS reaction in the most favourable case, two platinum wires are employed as catalytic electrodes, thus identifying the near-ideal case with a minimum overpotential of 1.73 V. For the 2-electrodes LSV measurements, a maximum potential difference of 3.5 V (Fig. 2a, continous line) is applied between the two electrodes. Similarly, for the two scenarios with SnO$_2$|CuSCN@P3HT and SnO$_2$|P3HT a LSV is performed. For the case

of SnO$_2$|CuSCN@P3HT, an onset potential of 3.03 V is identified as a threshold for the WS reaction to occur. On the other hand, the SnO$_2$|P3HT configuration displayed a low current (<0.5 mA cm$^{-2}$ @ 3.5 V) throughout the whole potential sweep with no effective current onset potential, suggesting that only negligible capacitive processes occurs rather than an actual WS reaction: this results is in agreement with literature data where P3HT is used as co-catalyst for the same purpose in combination with other active supports.[32] Thus, a clear difference in the electrochemical behaviour exists between the two examined systems, the first showing an effective onset potential at approximately 3 V, whereas the second fails to show any significant catalytic activity in the potential range swept, suggesting that its overpotential lies at least 0.5 V above the first one. The early overpotential in the composite HTM can be ascribed both to a positive interaction between CuSCN centers and P3HT chains and to a favourable mobility of holes in P3HT, having a lower energy barrier to cross for moving freely within the layer. Overall, these outcomes allows us to infer that the nano-engineered HTM employing CuSCN-NP additives effectively strongly reduces the potential required to perform the water oxydation reaction. This may be due to the most favourable energetic alignment of the valence band of CuSCN with respect to the water oxidation potential.[15] To the best of our knowledge, this is the first experimental evidence of a WS process involving P3HT and CuSCN.

To investigate further in detail the electrochemical behaviour of the HTM, the LSV measurements have been repeated under simulated sunlight (Fig. 2a, dotted lines). Pure P3HT does not display appreciably different behavior with respect to dark. Indeed, despite a slight increase in current density (50 µA cm$^{-2}$), the potential is still too low to promote the water splitting reaction. On the other hand, a substantial improvement can be detected when CuSCN@P3HT is employed. When illuminated, a further overpotential reduction ($\Delta V \approx 100$ mV) is achieved. Furthermore, at earlier potential than the onset, an increased capacitive current is recorded which may be attributed to the binding of oxygen on the polarized surface of the electrode. This photo-enhancement can be ascribed to a synergistic effect of P3HT and CuSCN. In this particular case, when photogenerated excitons in P3HT reach the interface with the CuSCN-NP, lower applied potential is needed for them to be split into free charges and holes to flow into the CuSCN. Here, thanks to the more favourable energetic alignment of the CuSCN valence band (VB) with respect to the oxygen evolution potential, holes injection to the electrolyte is favoured. The peak in the high potential region is attributed to the reduction of SnO$_2$ to Sn, according to the Pourbaix diagram.[33]

However, in a full device architecture, i.e. when the HTM are coupled to the perovskite layer, the photogeneration contribution of P3HT is negligible due to the shadowing of the main photoactive material. In that configuration, photogenerated holes coming from the perovskite layer flow in the CuSCN@P3HT HTM, where they can interact with environmental humidity and start binding to oxygen atoms in water molecules. The other half reaction is thought to occur on the perovskite interface, where photogenerated electrons with enough potential energy can reduce water to molecular hydrogen.[34]

In order to get deeper information on the charge transfer process happening within the composite material, we performed an electron paramagnetic resonance (EPR) analysis of the samples in dark conditions.

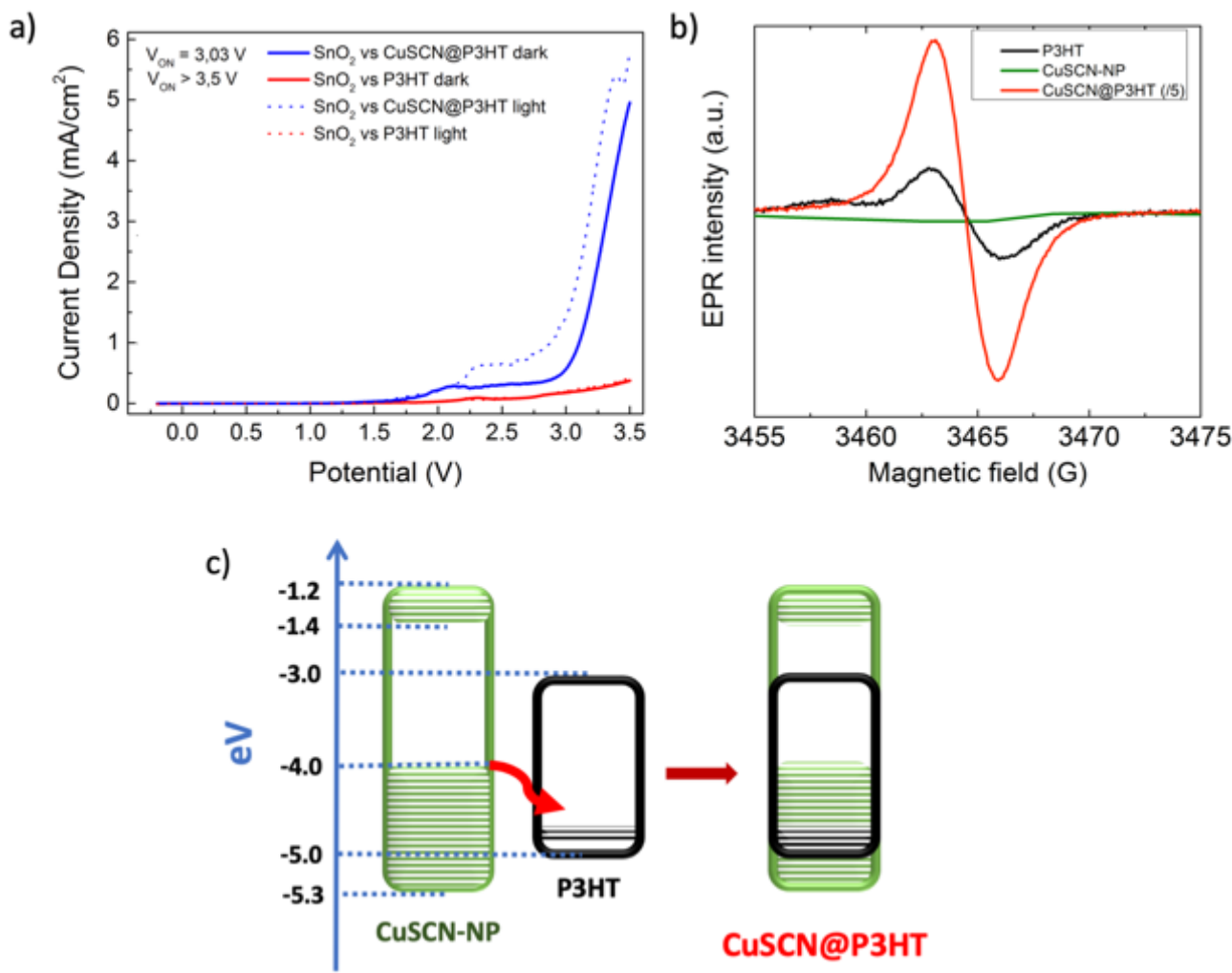

**Figure 2.** Electrochemical and physico-chemical characterization of the CuSCN@P3HT blend. a) Linear sweep voltammetries in dark (full lines) and in light (dotted lines) of two electrode systems based on CuSCN@P3HT/Au or pristine P3HT/Au anodes and SnO$_2$/ITO cathodes. b) EPR spectra in dark at low temperature of pristine P3HT, CuSCN-NP and CuSCN@P3HT. The amount of P3HT across the two different samples is the same. For the sake of clarity, the EPR signal for CuSCN@P3HT is reduced by a factor of 5. c) Energy levels diagram of CuSCN@P3HT (right), illustrating qualitatively the hole transfer process obtained after blending CuSCN-NP with P3HT (left). The P3HT work function value was taken from previous reports of some of us.[21,36] The dashed areas qualitatively represent filled states in CuSCN-NP (green) and in CuSCN@P3HT (black).

This technique is very sensitive to the presence of localized unpaired electrons within materials, therefore allowing us to better highlight the electrical role of the CuSCN-NP dispersed in the P3HT phase. In the present case, it reveals a considerable increase of the characteristic pristine P3HT signal for long-lived polarons, *i.e.* trapped positive charges, that naturally occur in *p*-type organic semiconductor material due to partial oxidation (Fig. 2b).[35] Such a result indicates that an overlap of the electronic bands of the two materials occurs, which may allow charge transfer between CuSCN and P3HT. This result is in agreement with an energy levels analysis conducted through Kelvin Probe (KP) microscopy and XPS. From KP, a value for the Fermi level of CuSCN-NP at -4 eV is obtained (Fig. S4a), whereas the XPS valence band spectrum, obtained employing the Mg k-alpha radiation, shows several bands which are in agreement with experimental and calculated spectra of CuSCN (Fig. 1c right bottom panel).[28] The on-set of the valence band photoemission, corresponding to the valence band maximum, is found at 1.3 eV below the Fermi level, thus allowing us to design an energy level diagram, as reported in Fig. 2c. The intimate contact between P3HT and CuSCN-NP that favors charge transfer (schematically sketched in Fig. 2c) is further proved by examining spin relaxation processes through EPR. Fig. S4b reports the decay of transverse magnetization over time for P3HT charged species. It is evident that the spin relaxation time for P3HT is significantly shortened when the polymer interacts with proximal Cu(I) centers, i.e. CuSCN acts as a "magnetic" quencher for P3HT charged states, in analogy with an optical quencher in a time-resolved photoluminescence experiment. In addition, the spatial range of the magnetic interactions leading to increased spin relaxation implies a close contact between the two species involved (less than a few nanometers).

To summarize, the spectroscopic and structural evidences suggest that the novel organic/inorganic nanocomposite CuSCN@P3HT features intriguing properties that perspect application in PSC. These are, namely, the platelet-like morphology of the CuSCN-NP, that is coherent with a vertical architecture of the PSC in which each layer is processed through spin coating, the ability of P3HT to efficiently disperse the inorganic fillers, the favourable overlap of energetic bands, which may favour a charge transfer process, and, more relevantly, the experimental evidence that proves the good WS activity of the nanocomposite compared to the pristine polymeric HTM. This allows the fabrication of the first rationally design PSC, to the best of our knowledge, that incorporates a WS active HTM.

**Perovskite Solar Cells Fabrication, Characterization and Stability Test**

PSC containing the binary blend HTM were fabricated as described thoroughly in the Experimental Section. P3HT and three CuSCN@P3HT HTM with two different concentrations of CuSCN-NP in the polymer matrix (5 and 10%, respectively) were employed. A preliminary screening of the HTM moisture compatibility was carried out before preparing PSC, through water contact angle measurements. Wettability characterization performed on the different P3HT-based HTM do not show any significant variation in the organic-inorganic composites compared to pristine P3HT (Fig. S5 reports the case of pure P3HT and of 10% CuSCN@P3HT, while the case of the intermediate CuSCN concentration is not shown since identical, with an approximate contact angle value of 100°) and thus we can infer that the well-known hydrophobic character of this polymeric material remains unchanged upon blending. Furthermore, inspection of the spin-casted layers underneath water drops reveals smooth films, which point out at a good morphological homogeinity of the CuSCN@P3HT composites. These evidences prospect a favorable use of the blends as good water-resistant HTM for direct architecture PSC.

The procedure previously described for preparing mixed-cation perovskites[37] is applied also here for the construction of the active layer. A standard PSC architecture based on ITO/SnO$_2$/(FAPbI$_3$)$_{0.87}$(CsPbBr$_3$)$_{0.13}$/HTM/Au is built for each device, containing one of the different HTM whose performance has to be compared. Devices are characterized morphologically through scanning electron microscopy (SEM). The good dispersion of the CuSCN-NP in the polymer film makes difficult to discern the presence of the inorganic nanofillers from top view images of the perovskite/HTM interface, even if the comparison with the case of pure P3HT highlights the existence of entities having a different contrast with sizes of hundreds nm in the composite HTM (Fig. S6a,b). However, the platelets are clearly visibile in the cross sectional image shown in Fig. S6d, in which horizontal structures (indicated with blue arrows) of several hundreds of nm lenght are detectable within the HTM. To make this layer more visible, the polishing step was on purpose avoided in the samples for cross-sections, thus allowing the better observation of the intrinsic HTM morphology, which appear more rigid in the composite compared to P3HT, where several pinholes are visible (indicated by yellow arrows in Fig. S6c).

In order to verify the impact of the WS process happening within the new composite HTM, the as-prepared PSC are freshly tested and then stored both in a glovebox ("dry" environment) and in a home-made container where a water-saturated atmosphere (RH> 80%) is generated ("humid" environment). The devices are maintained in the two different conditions for 28 days and photovoltaic data are acquired at different time points during the endurance test (see Fig. S7 in the SI for the overall data).

In Table 1 and Table 2 the summary of the figures of merit for all devices is reported at day 0 and day 28, both in dry and in humid. An "hysteresis factor" is calculated as the ratio between average

PCE values obtained from forward (f) and reverse (r) current density-voltage (J-V) scans: the higher is the value, the higher is the hysteresis.

Table 1. Average figures of merit for PSC measured as prepared (d 0) and after 28 days (d 28), stored in a water-saturated atmosphere with RH>80% ("humid" environment). Figures of merit analysis is performed only on reverse scans, for the sake of clarity. The hysteresis factor helps the reader to understand the overall photovoltaic results derived from both the forward and reverse scans.

| *Devices in humid* | *PCE* | | *ΔPCE (%)* | *Hysteresis Factor* | | *Jsc (mA cm$^{-2}$)* | | *Voc (V)* | | *FF* | |
|---|---|---|---|---|---|---|---|---|---|---|---|
| | *d 0* | *d 28* | *d28* | *d 0* | *d 28* | *d 0* | *d 28* | *d 0* | *d 28* | *d 0* | *d 28* |
| *P3HT* | 10.0 | 6.77 | -32.5 | 1.57 | 2.11 | 18.49 | 17.45 | 0.91 | 0.77 | 0.59 | 0.49 |
| *5% CuSCN@P3HT* | 8.95 | 7.91 | -11.6 | 1.96 | 1.57 | 19.12 | 17.11 | 0.87 | 0.90 | 0.54 | 0.50 |
| *10%CuSCN@P3HT* | 9.63 | 7.19 | -25.3 | 2.02 | 1.65 | 19.41 | 16.77 | 0.86 | 0.87 | 0.57 | 0.48 |

Table 2. Average figures of merit for PSC measured as prepared (d 0) and after 28 days (d 28), stored in a dry box ("dry" environment). Figures of merit analysis is performed only on reverse scans, for the sake of clarity. The hysteresis factor helps the reader to understand the overall photovoltaic results derived from both the forward and reverse scans.

| *Devices in dry* | *PCE* | | *ΔPCE (%)* | *Hysteresis Factor* | | *Jsc (mA cm$^{-2}$)* | | *Voc (V)* | | *FF* | |
|---|---|---|---|---|---|---|---|---|---|---|---|
| | *d 0* | *d 28* | *d28* | *d 0* | *d 28* | *d 0* | *d 28* | *d 0* | *d 28* | *d 0* | *d 28* |
| *P3HT* | 10.04 | 10.3 | -0.1 | 1.00 | 1.69 | 20.16 | 18.94 | 0.90 | 0.92 | 0.54 | 0.59 |
| *5% CuSCN@P3HT* | 8.95 | 7.58 | -15.3 | 2.05 | 1.68 | 18.32 | 18.16 | 0.87 | 0.81 | 0.56 | 0.50 |
| *10% CuSCN@P3HT* | 9.63 | 6.56 | -31.9 | 2.06 | 1.63 | 18.24 | 17.60 | 0.87 | 0.78 | 0.56 | 0.47 |

Analysis of the data reported in Tables 1 and 2 shows that the reference PSC based on the pristine P3HT HTM undergo a degradation process after almost one month storage in humid conditions that is not detected in devices kept in a dry environment (32.5% loss in average PCE against 0.1% loss), whereas the performance of the PSC containing the additional 5 wt% CuSCN-NP contribution in a water-saturated atmosphere changes very moderately (-11.6%). In this last case, the hysteris factor is even slightly improved (from 1.96 at d 0 to 1.57 at d 28). Generally humidity affects the perovskite layer integrity (and more specifically the corresponding device short circuit current density, $J_{sc}$), therefore a wettability determination would be mandatory to exclude an additional hydrophobic effect of the CuSCN additives that protects the photo-active layer in an extremely humid environment. As discussed previously, the presence of the CuSCN-NP in P3HT up to 10 wt% does not change the water contact angle with respect to the already very hydrophobic P3HT (see Fig. S5). Therefore, assuming that moisture can penetrate in all devices, other effects must justify the impressively stable performance of the 5wt% CuSCN@P3HT containing devices stored in humid conditions, especially of the lowest hysteresis and the highest open circuit voltage (Voc) compared to all other cases examined in the same conditions.

In Figure 3 PCE and Voc statistical data for a large number of as-prepared (d 0) and aged (d 28) devices based on P3HT and 5% CuSCN@P3HT HTM are plotted, in both forward and reverse scans. As mentioned above, the average PCE for the P3HT-based PSC is notably decreased in humid atmosphere, whereas it remains almost unvaried for the 5% CuSCN@P3HT-based ones. A best pixel with 9.6% PCE is found for the latter at d28, whereas a 7.85% PCE results for the reference (reverse scans in Fig. 3a and 3b). In addition, the dispersion of PCE data is significantly high for P3HT-based devices at d 28 in humid conditions, whereas it is much smaller for the CuSCN-contanining PSC. This aspect points out at the fact that moisture must have a relevant role in determining the fate of the solar cell. However, knowning that water tends to degrade the perovskite layer affecting Jsc values (see Fig. S8a,b in the SI) and, consequently PCE, the analysis of the other figures of merit, Voc (Fig. 3c,d) and fill factor (FF, Fig. S8c,d), emerges as an helpful tool to understand more in detail the nature of the degradation mechanism.

Voc of solar cells is analyzed in Fig. 3c and Fig. 3d: Voc values for P3HT-only devices are strongly affected by the high humidity, while they are slightly changed or even improved when CuSCN is present. The opposite happens in dry conditions: Voc (and also Jsc) is improved in the P3HT control solar cell, while a serious degradation occurs with CuSCN-NP. Both high Voc and Jsc values are normally related to optimal interfaces between adjacent layers: at d 0, the energy alignment of the perovskite layer and the composite HTM is not ideal, being the work function of the latter very similar to the conduction band of the former (4 eV vs 3.39 eV). This may depress photocurrent and photovoltage, due to recombination of excitons. We can therefore infer that a dry environment is the sub-optimal working condition for our CuSCN@P3HT composite HTM in a standard architecture PSC, while the material requires a humid atmosphere to express its potential in stabilizing solar cells performance.

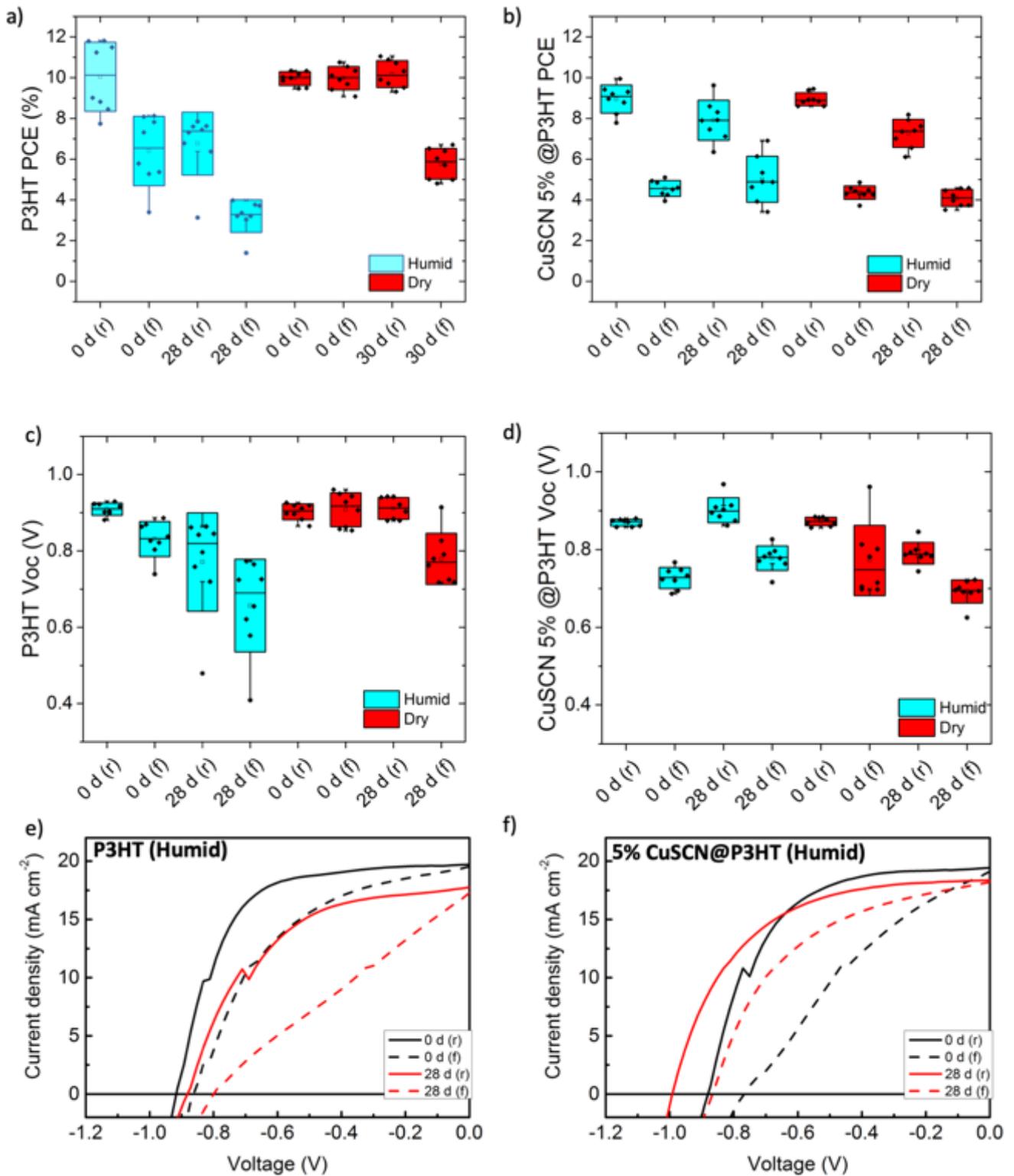

**Fig. 3.** Summary of PSC performance. Statistical data on PCE and $V_{OC}$ variations during 28 days in a water-saturated atmosphere (humid, pale blue) and in a dry atmosphere (dry, red) for PSC containing a,c) the reference P3HT and b,d) the 5% CuSCN@P3HT HTM. Best pixel J-V curves for the P3HT-based device e) and for 5% CuSCN@P3HT-based device f) as prepared (0 d) and after the 28 days of endurance test in high-humidity conditions (28 d). The complete dataset of photovoltaic parameters is reported in Fig. S7.

In this context, a spectroscopical characterization such as time-resolved photoluminescence (TRPL), that probes the ageing effect of the two opposite humidity conditions directly on the photoactive perovskite layer, can be helpful in better understand the degradation processes happening in the devices. We examined and compared the decay of the photoluminescence (PL) emission at 780 nm, to extract the lifetime of the carriers in the perovskite layers (see Fig. S9a,b in the SI). After the photoexcitation (600 nm) of the perovskite films in solar cell architectures, many processes may determine the overall lifetime of the carriers, among which most significant are carrier trapping, carrier extraction, and radiative recombination. It is worth to note that both carrier trapping and extraction in and from the perovskite layer are expected to decrease the carrier lifetime, while only the carrier extraction contributes to the photovoltaic efficiency. To get a qualitative insight into the degradation process, we compared the average lifetime of the carriers with the recorded average PCE of the solar cells (Figure 4a,b). The P3HT-only device aged in "dry" conditions displays a slightly shorter average lifetime ($<\tau>_{P3HT} = 10 \pm 1\ ns$) despite showing the higher photovoltaic performance. On the other hand, the 5% CuSCN@P3HT device reports a longer lifetime ($<\tau>_{CuSCN\ 5\%} = 11 \pm 1\ ns$) while reporting lower PCE. Therefore, this qualitative anticorrelation of lifetimes and PCE during the dry ageing suggest that different degrees of degradation in the charge extraction layer (HTM) are partly responsible for the losses in the photovoltaic efficiency. This translate in poorer carrier extraction and thus longer PL lifetimes. On the other hand, the qualitative comparison shows the opposite trend for ageing in "humid" conditions: over the series, shorter average lifetimes of carriers in the perovskite films are associated with poorer photovoltaic performances. Therefore, the comparison suggests that the humidity-driven degradation of the perovskite layer (i.e., increased carrier trapping) is responsible for the degradation of photovoltaic performances. Notably, these results suggest that in "humid" conditions the presence of 5% CuSCN-NP additives preserves the photovoltaic performances, i.e. preventing the perovskite degradation as quantitively demonstrated by fitting the PL decays.

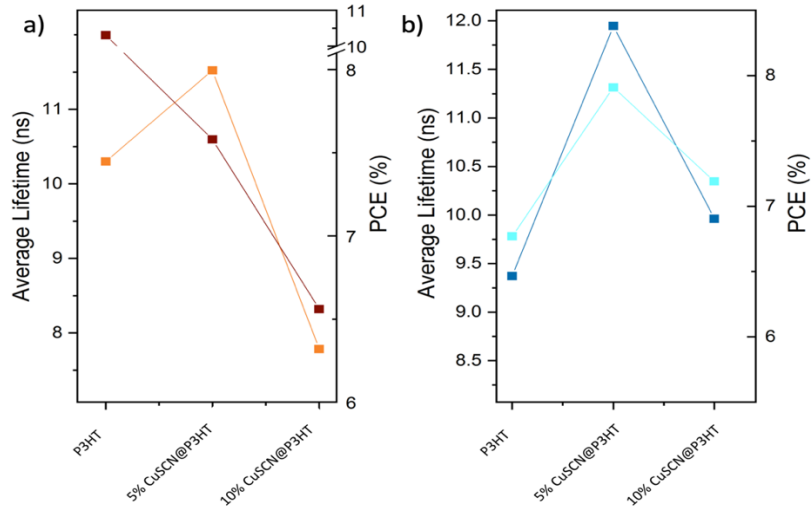

**Fig. 4.** Correlation between average PL lifetimes and PCE for "dry" a) and "humid" b) storage, respectively.

We measured excitation power dependent TRPL spectra (Figure S10 in the SI) and analyzed the kinetics of the PL decay according to the differential rate law:

$-\frac{dn}{dt} = k_1 n + k_2 n^2.$

Given the free carrier nature of hybrid lead iodide perovskites, the monomolecular recombination rate $k_1$ is ascribed to carrier trapping processes, while the bimolecular recombination rate constant is ascribed to the radiative recombination. However, since we studied the carrier dynamics in solar cell architectures, $k_1$ also contains information on the carrier extraction and therefore can be expressed as $k_1 = k_T + c_{ext}$. As reported in Table S1 in the SI, $k_1$ values in the humid conditions are lower for the 5% CuSCN@P3HT cell and, considering PCE data, we can safely determine that the presence of small quantities of the inorganic semiconductor additive in the HTM help in protecting the perovskite layer.

In order to better understand the origin of this peculiar behavior of the 5% CuSCN@P3HT composite HTM observed in humid conditions and in which extent it can be related to the WS process hyphotetized, we tracked the WF of the different HTM in the two opposite environmental conditions (humid and dry) using KP. Indeed, a better hole-extraction is usually linked to a favorable energy alignment between the HTM and the top electrode (Au in this case). A *pn* junction is generated by spin coating the different *p*-type HTM semiconductors on top of FTO substrates, that in this case act as the *n*-type semiconductor. In such a way, a built-in potential is established within the devices, that simulates the cell working mechanism during Voc application, *i.e.* a flow of charges towards FTO (the negative ones) and through the HTM (the positive ones). The estimation of the WF can be reasonably associated to the actual energy of the Fermi level at the FTO/HTM interface and its eventual shifting is strongly related to the presence of active defects in the band gap of the semiconductors that increase energy losses (*i.e.* decrease Voc). In Fig. 5a the average measured WF of the different HTM is reported. In order to ease the correct understanding of KP data, the HTM WF values are reported as a difference with respect to the gold WF. Due to the nature of the tip used for the measurement (see SI for details) this choice is particularly helpful, because Au represents also the anode metal at which holes are collected in the PSC. Therefore, the more positive is the energy difference, the higher is the WF of the HTM and the better the HTM should work in a PSC as transport layer in a given environmental condition. P3HT and 5% CuSCN@P3HT clearly show the most suited energetic conditions with an actual WF of about -5.1 eV (almost no difference with respect to Au, that we assign to -5.1 eV, as from literature data[38]). On the contrary the WF of 10% CuSCN@P3HT is located at -5 eV. In dry conditions the system behaves completely different, with P3HT having a WF of about -4.8 eV, and thus confirming the undoped nature of the pristine semiconducting polymer.[18] 5% CuSCN@P3HT shows large data dispersion, a result that is in accordance with EPR results in dark (Fig. 2b) that, by confirming the charge transfer process between the two materials in the binary composite, suggests how the WF could change locally within the HTM depending on the presence or not of CuSCN-NP in a given position.

From the comparison, we can state that humid conditions favor an increase in the WF of the different HTM containing CuSCN (pail blue pattern), *i.e.* we can attest that a *p*-doping process occurs. Considering the WS process, we can assume that the doping agent for P3HT is molecular oxygen derived from the water oxidation reaction. Based on this, it derives that the WS process occurs simultaneously to PSC operation, as described in Fig. 5b. To the main PSC working mechanism (i.e. the flow of charges into the external circuit), a parasitic mechanism (i.e. the WS itself) is added to the overall PSC operation: when water diffuses within the top layers, some photoexcited holes immediately convert it to molecular oxygen, that, due to the very short distance between CuSCN-NP centers and P3HT, *p*-dopes the semiconducting polymer (i.e. the HTM Fermi level is lowered towards optimal transport conditions). On the other hand, some photoexcited electrons in the conduction band of the perovskite are employed to reduce water to molecular hydrogen. In this way, water molecules can be *in situ* degraded, while perovskite layer integrity and, consequently, PSC operation can be preserved from undesired chemical degradation.

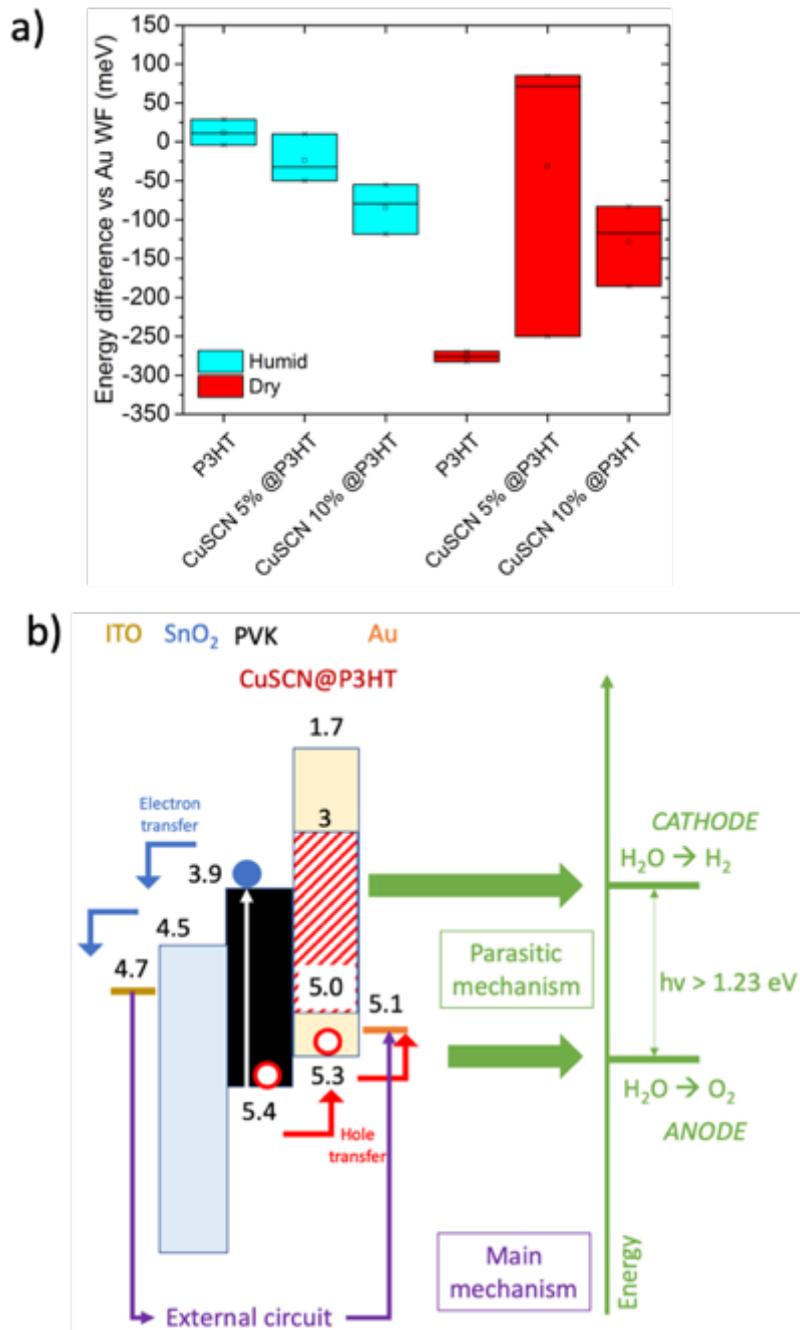

**Fig. 5.** a) Chart summarizing Kelvin probe analysis: "humid" and "dry" behaviour are compared in different colours. The zero value in the y axis represent the WF of Au. b) Proposed mechanisms for the two processes happening within the WS-integrating PSC. The "main mechanism" (violet) is the regular working process of the PSC. The "parasitic mechanism" is the WS-process occurring at the perovskite/CuSCN@P3HT interface. This last one takes place because, during the endurance test in a water-saturated atmosphere, moisture, diffusing from the top of the device, is locally oxidized by the CuSCN-NP center contained in the composite HTM, forming *in situ* molecular oxygen, that p-dopes the semiconducting P3HT matrix. The required energy difference for WS is guaranteed by the valence band position of CuSCN and conduction band position of the perovskite (here indicated as PVK) acting as, respectively, anode and cathode sites for the WS reaction, whereas the flow of the charges is supported by the built-in potential (in dark during the storage) or by the bias (during solar cell electrical characterization).

Within this scenario, we can state that humid conditions favor the operational activity in a CuSCN-doped HTM but, at the same time, a water-saturated atmosphere degrades PSC. It is thus necessary to understand what is the origin of the *p*-doping and why the PSC containing 5% CuSCN@P3HT does not degrade significantly in high humidity conditions. We suggest that behind the observed phenomena a WS-process in the presence of CuSCN-NP is taking place, as sketched in Fig. 5b. It is known indeed that such an inorganic semiconductor is used for improving the yield of the oxygen evolution reaction in combination with metal oxides, due to its wide-bandgap, the good band alignment with the water oxidation potential and the good hole conductivity.[15,39,40] We speculate therefore that, in the present case, water molecules diffusing from the top of the unsealed PSC into the composite HTM during the endurance test carried out in a water-saturated atmosphere, react with the CuSCN-NP water oxidation centers when the device is biased during J-V scans, by forming molecular oxygen. On the other hand, the perovskite layer acts as center for water reduction to molecular hydrogen. It is well known that P3HT can be *p*-doped in presence of oxygen and light.[41–43] Therefore, in the binary HTM system, in which an unfavorable starting situation, deriving from a misalignment of the energy levels, depresses PCE, the rise in oxygen content following water and light exposure, allows the *p*-doping of the polymeric hole conductor and thus pins down the Fermi level towards an optimal value, by *in-situ* reducing energy losses. The opposite happens when moisture is not present: in this case the polymer doping is not effective and only a charge transfer process takes place between the constituents of the HTM, thus locally altering the Fermi level of the system in an chameleonic way, that depends on the specific point in the HTM where CuSCN-NP are either present or not.

The implementation of a WS-integrated PSC in high humidity environments has an impact on the figures of merit: i) a robust increase of the average Voc (see Fig. 3d); ii) a decrease of the electrical hysteresis (see Table1) due to the lower amount of water reaching the perovskite, being the former transformed into oxygen (comparable Jsc, see Table 1), and iii) a slightly higher fill factor (FF) due to a reduced series resistance allowed by a better quality perovskite/P3HT-based HTM interface (see Fig. S6).

**Conclusions**

In this work, for the first time, the water oxidation capability of an inorganic semiconductor is exploited as a tool to preserve the functionality of a lead halide perovskite-based solar device maintained under significant stress-conditions of high relative humidity. This concept is *per se* relevant for opening up new perspectives in the stabilization of last generation optoelectronic devices based on water-sensitive species.

The active role of the wide-band gap CuSCN-NP in the composite HTM that we described here is not only that of electron-blocking species (which is anyway relevant, as seen from the high Voc of the solar cells after one month of moisture exposition), but also of selective oxidants for the moisture diffusing within the HTM during PSC ageing. In this way, the oxygen production within the HTM triggers the favorable *in situ p*-doping of P3HT.

The here presented results, specific on PSC, interestingly pioneer a research field in which also other inorganic nanomaterials besides CuSCN-NP could be employed as mediators for promoting *p*-doping of semiconducting polymers-based optoelectronic devices operating in humid environments, after designing and applying a proper tuning of the adjacent layers and of the overall energy levels landscape.

**Experimental Section**

*Materials*. All materials and solvents were purchased from Sigma Aldrich, unless otherwise specified, and used as received. $PbI_2$, $PbBr_2$ and CsBr were purchased from Alfa Aesar. Formamidinium iodide (FAI) was obtained from Dyesol. P3HT was purchased from Merck (Mw = 94.100 g mol−1, PD = 1.9, RR = 95.5%).

*Synthesis of CuSCN-NP*. The material was produced through CFHS in a lab-scale reactor. The reactor was made from off-the-shelf Swagelok™ parts and equipped with two patented confined jet mixers.[44] The mixers were fed by three Primeroyal K diaphragm pumps (Milton Roy, Pont-Saint-Pierre, France, pressurized to 24.1 MPa; see Figure S1 in the Supporting Information –S.I.- for a sketch of the reactor). Pump 1 was employed for the supercritical water line at 80 ml min$^{-1}$, where water was heated in line at 450 °C using a custom-made 7 kW heater. Pump 2 was used to feed a 0.5 M thiourea solution to the first mixer at 40 ml min$^{-1}$, to allow its thermal decomposition leading to the production *in-situ* of SCN- at 380 °C.[45] Pump 3 was then used to pump a 0.1 M $Cu(NO_3)_2 \cdot 3H_2O$ solution to the second mixer at 100 ml/min, allowing the precipitation of CuSCN from the mixing with the SCN- solution at 260 °C. The residence time between the first and the second mixer was 2.28 s. The product was collected as slurry in a simple flask. Once the solid product settled, the clear solution above was carefully removed. The product was isolated though centrifugation (10 min at 10 000 rpm), then it was washed by adding fresh water and isolated again though centrifugation (10 min at 10 000 rpm) for four times. The clean solid was freeze-dried, grinded and stored in a common Falcon tube.

*XRD analysis.* XRD measurements for crystal phase identification were carried out at room temperature by a Bruker D8 Advance Diffractometer equipped with a Göbel mirror, using the Cu-Kα radiation. The angular accuracy was 0.0010° and the angular resolution was better than 0.01°.

*XPS analysis.* CuSCN-NP powder was ground in a mortar and pressed against an indium pellet on a sample holder. The sample was analysed using a Kratos Axis Ultra DLD apparatus at a base pressure of $10^{-9}$ mbar, using a Mg k-alpha X-ray source operating at 15 kV accellerating voltage and 5 mA emission current. Charge neutralization was used during the measurement. The wide spectrum was acquired with a fixed pass energy (PE) of 160 eV and step size of 1 eV, while high resolution spectra of the Cu 2p, Cu LMM, O 1s, N 1s, C 1s and S 2p regions were measured at PE = 10 eV and step size of 0.1 eV. The valence band region was measured with PE = 10eV and step size 25 meV. The binding energy (BE) was calibrated assigning BE = 285 eV to the aliphatic C 1s, which results in an estimated uncertainty of 0.2 eV on absolute energy values. Spectra were analysed using CasaXPS software (version 2.3.19)

*SEM and EDS analysis.* As synthesized CuSCN-NP were gently attached to conductive carbon tape. After carefully removing the excess, particles were sputter coated with a 10 nm Au layer for imaging and with a carbon layer for EDS analysis. Scanning electron microscopy was performed on a JEOL JSM-6490LA SEM operating in high vacuum with a beam acceleration of 15 kV.

*TEM, DLS and UV-vis-NIR analysis.* A 1 mg ml$^{-1}$ suspension of CuSCN-NP in water was obtained by dispersing the powder with the aid of tip ultrasonication (for 20 s at 40 % amplitude) and a surfactant. Polyvinylpyrrolidone (PVP, 1.3 MDa) and keratin (the protein extracted from wool, using a previously described protocol)[46] were used as surfactant. The ratio between CuSCN-NP and surfactant was kept as low as possible to obtain stable suspensions, namely at 10:1 in weight. After letting the suspension settle overnight to remove large particles aggregates, 10 µL were dispensed on a 150 mesh Cu grid coated with an ultrathin carbon film that was plasma treated to increase hydrophilicity. Extra solution was wicked off from the side using filter paper and grids were let to dry overnight before imaging. Two surfactants were used to be able to exclude possible artifacts due to surfactant self assembly. TEM images were recorded on a JEOL JEM-1011 apparatus operating at 100 kV. The same suspensions were used for DLS analysis after further dilution at 0.01 mg ml$^{-1}$. The DLS spectrum was recorded on a Zetasizer Nano S (Malvern Instruments) at 20 °C. A quartz cuvette was employed with 1 cm optical path. Optical properties of CuSCN (n = 2.2, k = 0.02) used in the modeling, were assumed from literature reference.[47] UV-vis-NIR absorption spectra of the CuSCN-NP suspensions used for Tauc plot calculation were recorded on a Cary5000 Varian spectrometer.

*Preparation of CuSCN@P3HT blends.* P3HT was dissolved in chlorobenzene at a concentration of 15 mg mL$^{-1}$ and the resulting solution was stirred overnight at 70 °C on a hot plate. CuSCN-NP (3.8 mg, 7.5 mg and 11.3 mg for, respectively, 5, 10 and 15 wt% blends, with respect to the P3HT weight) were added each to separated 5 mL portions of the mother P3HT solution and the resulting mixtures were sonicated for 120 s with a tip-sonicator operated at 280 W using a 50% pulse scheme.

*Electrochemical characterization.* The electrochemical tests were performed in a 2-electrodes configuration using an Autolab PGSTAT302N potentiostat/galvanostat station, equipped with Nova 2.1 (Metrohm) software package. Measurement were performed in a pH = 7, 0.1 M Phosphate buffer solution. A vinyl tape mask has been employed to define an electrochemical active area of 0.2 cm$^2$. Linear Sweep Voltammetry (LSV) has been employed to determine the electrochemical response of the electrodes for a potential range going from 0 V to 3.5 V with a scan rate of 10 mV/s. The electrodes have been prepared according to the material processing described in the PSC fabrication section. For the P3HT-based electrode the deposition order has been inverted, first depositing gold on a ITO substrate and then spin coating the pure polymer or the CuSCN@P3HT blend. Measurements have been performed also under simulated sunlight by means of A 300 W Xenon light source (Lot Quantum Design, model LS0306), equipped with AM 1.5 G filters. In both the tested conditions (pure P3HT and CuSCN@P3HT), light has been shone passing through the glass/ITO/SnO$_2$ electrode first.

*EPR measurements.* EPR measurements were performed on a Bruker ELEXSYS E580 spectrometer at X-band (9-10 GHz) mounting an ER4118X-MD5 dielectric cavity. Measurements were carried out in dark either at room temperature or at 80 K, employing a liquid-N$_2$ cooling system. Continuous-wave spectra were typically acquired at 20-40 μW microwave power and 0.8 G field modulation amplitude. Spin relaxation experiments were performed at 80 K using a standard echo-decay pulse sequence π/2 – t – π – t – det (pulse length 16 ns and 32 ns for π/2 and π pulses, respectively). Echo decays starting at t0 = 140 ns were acquired at the magnetic field corresponding to maximum absorbance and involved echo integration and 2-step phase cycling. All samples were measured under vacuum (~10-5 bar) to avoid disturbances due to paramagnetic O$_2$. In the case of the samples containing P3HT, fixed volumes of the corresponding solutions were inserted into quartz tubes and evaporated under vacuum, followed by tube sealing.

*PL measurements.* For time-resolved photoluminescence (TRPL) measurements, the excitation was provided by Coherent Libra regenerative amplifier (50 fs, 1 KHz, 800 nm) seeded by a Coherent Vitesse oscillator (50 fs, 80 MHz). From the fundamental output, 600 nm laser pulses were generated using a Coherent OPerA-Solo optical parametric amplifier. The emission from the samples was collected in backscattering geometry (150° collection angle), and focused by a pair of lenses into a spectrometer. TRPL was collected using an Optronis Optoscope streak camera system, with a 10 ps temporal resolution.

*PSC fabrication and characterization.* ITO-coated glass was etched with zinc powder and 2M aqueous HCl solution for electrode pattern. The ITO substrates were washed with 2% Hellmanex in water, deionized water, iso-propanol, acetone, and iso-propanol sequentially in a sonication bath for ~15 min, followed by $O_2$ plasma cleaning for 10 min. A patterned and cleaned ITO substrate was covered with a ~10 nm thick $SnO_2$ layer by spin-coating of a diluted $SnO_2$ nanoparticle solution (Alfa Aesar) and annealed at 180 °C for 1 h. The mixed cation FA/Cs perovskite was prepared by dissolving different molar quantities of 1.21 M $PbI_2$, 1.21 M FAI, 0.24 M $PbBr_2$, and 0.24 M CsBr in 1 L of N, N-Dimethylformamide (DMF) and dimethylsulfoxide (DMSO) at a volume ratio of 4:1. After stirring for 6 h at room temperature, the perovskite solutions became clear, and they were spin-coated with two steps (the first step at 1000 rpm for 10 s and the second step at 4000 rpm for 30 s). During the second step, 200 μL of anhydrous toluene was quickly dripped at the $15^{th}$ s. The thin films were then transferred to a hotplate and annealed at 170 °C for 10 min. The whole synthesis process was conducted in the nitrogen-filled glovebox. P3HT-based HTM were deposited on the perovskite film by spin-coating at 2000 rpm. This fabrication process was carried out under controlled atmosphere in a glove-box. Finally, a 75 nm thick layer of gold was thermally evaporated on top of the device at a pressure of $1 \times 10^{-6}$ mbar to form the back contact. The active area of the complete device results 0.0935 $cm^{-2}$. SEM images of analogous devices lacking the top metal electrode in top-view and cross-section were obtained on a Zeiss Sigma HD field-emission SEM, working at 5 kV. J-V characteristics were measured with a computer-controlled Keithley 2420 source meter in air without any previous device encapsulation. The simulated Air Mass 1.5 Global (AM 1.5G) irradiance was provided with a class AAA Newport solar simulator. The light intensity was calibrated with a silicon reference cell with a spectral mismatch factor of 0.99. Scan rates were 0.2 V $s^{-1}$. The forward scan started from 0 V (the short-circuit condition) to 1.4 V, while reverse scan from 1.4 V to 0 V. The pre-conditional stress was not applied for PV measurements. For degradation test, RH was controlled in a closed glass jar where water was vaporized from a saturated salt solution of potassium chloride. PSC were kept in the glass jar under a RH of 88 ± 1% in dark at room temperature.

*Kelvin probe measurements.* Kelvin probe microscopy measurements on CuSCN-NP were carried out in ambient air with an atomic force microscope MFP-3D by Asylum Research (Oxford Instruments, UK). A probe AYELEC.02 was used, with nominal cantilever resonance frequency of 300 kHz, and tip coated with 25 nm thick Ti/Ir film. The scan was two-pass, with surface potential measured during the second pass at elevated height (tipically 100 nm). The work function of the tip was first determined, by measuring V on highly-oriented pyrolitic graphite (HOPG), whose work function was assumed to be 4.40 eV. Measurements were carried out both on sample powder, spread on a Silicon wafer, and on a thick (bulk) layer of sample powder, pressed to form a solid pellet, finding consistent values, within the uncertainty (±100 mV). The obtained value resulted from averaging the means from N=4 images of typical 5 μm scan size, in different sample locations. Kelvin probe measurements on solar cells were carried out on a KP Technology single point system, equipped with a 2 mm gold coated tip on samples stored for 28 d in humid (RH = 88.1%) and dry conditions. Each measurement was left to stabilize for 5 min after which the measured work function was regarded as stable. The samples for KP were prepared by spin-coating the different HTM directly onto FTO substrates following the same procedure used for depositing them on top of the perovskite layer during PSC fabrication.

**Acknowledgments**

F.L., S.G. and G.M. acknowledge the "Centro Studi di Economia e Tecnica dell'Energia Giorgio Levi Cases" of the University of Padova through the AMON-RA project for financial support. M.K. acknowledges funding from EU Horizon 2020 via a Marie Sklodowska Curie Fellowship (Project No. 797546). T.G. and S.G. thank the DFG for financial support via the GRK (Research Training Group) 2204 "Substitute Materials for Sustainable Energy Technologies". Prof. Jawwad Darr (UCL, London) is gratefully acknowledged for the possibility to perform the CuSCN synthesis on his continuous flow hydrothermal synthesis reactor. We sincerely thank Dr. Marcello Righetto and Prof. Tze-Chien Sum of the National University of Singapore for time-resolved photoluminescence measurements and analysis. We also thank Dr. Peter Topolovsek (CNST@IIT Milano) for the help with KP measurements on pn junctions and Prof. Lorenzo Franco's group (University of Padova, Dept. of Chemical Sciences) for EPR measurements. We sincerely acknowledge Dr. James Ball for the insightful discussion and suggestions.

**Author contributions**

F.L. and T.G conceived the idea of the work, designed and managed the project and the experiments. N.D. prepared the CuSCN-NP and performed XRD and UV-Visible characterization of the CuSCN-NP. G.P. characterized the CuSCN-NP with SEM-EDX, TEM and DLS. M.S. and M.P. performed XPS measurements on CuSCN-NP. M.Sa. performed KP measurments on CuSCN-NP. N.D. and M.K. prepared the P3HT blends. A.A. and A.M. designed the experiment and performed the WS splitting test. R.S. performed wettability measurments. M.K. fabricated the solar cells and performed the endurance tests. M.K. performed KP characterization. F.L. and T.G. analyzed the data. F.L. and T.G. wrote the manuscript. S.G., G.M., F.D.F. and A.P. revised the manuscript. All authors contributed to finalizing the draft and gave their final assessment to the submitted paper.

**Conflit of interests**

The authors declare no competing interests.